\documentclass[12pt,twoside]{article}
\usepackage{psfig}
\usepackage{amssymb}
\begin{document}
\begin{center}
{\Large\bf
FITTING THE SUPERNOVA TYPE Ia DATA\\[5PT]
WITH THE CHAPLYGIN GAS\\[5PT]}
\medskip
{\bf
J. C. Fabris\footnote{e-mail: fabris@cce.ufes.br},
S.V.B. Gon\c{c}alves\footnote{e-mail: sergio@cce.ufes.br} and P.E. de Souza\footnote{e-mail: patricia.ilus@bol.com.br}}\\  \medskip
Departamento de F\'{\i}sica, Universidade Federal do Esp\'{\i}rito Santo,
CEP29060-900, Vit\'oria, Esp\'{\i}rito Santo, Brazil \medskip
\end{center}
\begin{abstract}
The supernova type Ia observational data are fitted using a model with cold dark matter
and the Chaplygin gas. The Chaplygin gas, which is characterized by a negative pressure
varying with the inverse of density, represents in this model the dark energy responsible
for the accelaration of the universe. The fitting depends essentially on four parameters:
the Hubble constant, the velocity of sound of the Chaplygin gas and the fraction density of the
Chaplygin gas and the cold dark matter. The best fitting model is obtained with $H_0 = 65\,km/Mpc.s$,
$c_s^2 \sim 0.92\,c$ and $\Omega_{c0} = 1$, $\Omega_{m0} = 0$, that is,
a universe completely
dominated by the Chaplygin gas. This reinforces the possibility that the Chaplygin gas
may unify dark matter and dark energy, as it has already been claimed in the literature.
\vspace{0.7cm}
\newline
PACS number(s): 98.80.Bp, 98.65.Dx, 98.80.Es
\end{abstract}
The combined data from the measurements of the spectrum of anisotropies of the cosmic
microwave background radiation \cite{charles} and from the observations of high redshift supernova type
Ia \cite{riess,perlmutter} indicate that the matter content of the universe today may be very probabily described
by cold dark matter and dark energy, in a proportion such that $\Omega_{dm} \sim 0.3$ and
$\Omega_{de} \sim 0.7$. The distinction between cold dark matter and dark energy lies on the
fact that both manifest themselves through their gravitational effects only and,
at the same
time, on the fact that a fraction of this dark matter agglomerates at small scales (cold
dark matter) while the other fraction seems to be a smooth component (dark energy). The dark
energy must exhibit negative pressure, since it would be the responsable for the
present acceleration of the universe, as induced by the supernova type Ia observations,
while the cold matter must have zero (or almost zero) pressure, in order that it can
gravitationally collapse at small scales.
\par
The nature of these mysterious matter components of the universe is the object of many
debates. The cold dark matter may be, for example, axions which result from the
symmetry breaking process of Grand Unified Theories in the very early universe. But,
since the Grand Unified Theories, and their supersymmetrical versions, remain a theoretical proposal, the nature of cold dark
matter is an open issue.
\par
A cosmological constant is, in principle, the most natural
candidate to describe the dark energy. It contributes with a homogeneous, constant energy
density, its fluctuation being strictly zero. However, if the origin of the cosmological constant
is the vacuum energy, there is a discrepancy of about $120$ orders of magnitude between its
theoretical value and the observed value of dark energy \cite{carroll}.
This situation can be ameliorate, but
not solved, if sypersymmetry is taken into account. Another candidate to represent dark energy
is quintessence, which considers a self-interacting scalar field, which interpolates a radiative
era and a vacuum dominated era \cite{steinhardt,sahni,brax}. But the quintessence program suffers from fine tuning of
microphysical paremeters.
\par
Recently, an alternative to both the cosmological constant and to quintessence to describe
dark energy has been proposed: the Chaplygin gas \cite{pasquier,patricia,bilic,bento}.
The Chaplygin gas is characterized by the equation of state
\begin{equation}
\label{eoe}
p = - \frac{A}{\rho} \quad ,
\end{equation}
where $A$ is a constant. Hence, the pressure is negative while the sound velocity is
positive, avoiding instability problems at small scales \cite{jerome}.
The Chaplygin gas has been firstly been conceived in studies of adiabatic fluids
\cite{chaplygin},
but recently it has been identified in interesting connection with string theories
\cite{ogawa}.
In fact, considering a $d$-brane in a $d+2$ dimensional space-time, the introduction
of light-cone variables in the resulting Nambu-Goto action leads to the action
of a newtonian fluid with the equation of state (\ref{eoe}) \cite{jackiw}. The symmetries of this
"newtonian" fluid have the same dimension as the Poincar\'e group, revealing that
the  relativistic symmetries are somehow hidden in the equation of state
(\ref{eoe}).
\par
Considering a relativistic fluid with the equation of state (\ref{eoe}), the
equations for the energy-momentum conservation relations leads, in the case of a homogeneous and
isotropic universe, to the following relation between the fluid density and
the scale factor $a$:
\begin{equation}
\rho = \sqrt{A + \frac{B}{a^6}} \quad ,
\end{equation}
where $B$ is an integration constant.
This relation shows that initially the Chaplygin gas behaves as a pressureless
fluid, acquiring later a behaviour similar to a cosmological constant.
So, it interpolates a non-accelerated phase of expansion to an accelerate one,
in a way close to that of the quintessence program.
\par
In this work, we will constrain the parameters associated with the Chaplygin
gas using the supernova type Ia data. Specifically, we will consider a model
where the dynamics of the universe is driven by pressureless matter and
by the Chaplygin gas. The luminosity distance for this configuration is
evaluated, from which a relation between the magnitude and the redshift
$z$ is established. The observational data are then considered, and they are
fitted using four free parameters: the density fraction, with respect to the
critical density $\rho_c$, today of the pressureless matter and of the Chaplygin gas,
$\Omega_{m_0}$ and $\Omega_{c0}$ respectively, the sound velocity of the Chaplygin
gas today $\bar A$, in terms of the velocity of light,
and the Hublle parameter $H_0$. The sound velocity of the
Chaplygin gas today is given by
\begin{equation}
\bar A = \frac{A}{\rho_{c0}^2} \quad .
\end{equation}
It will be verified that the best fitting is obtained when
$\bar A \sim 0.92$, $\Omega_{m0} = 0$, $\Omega_{c0} = 1$ and $H_0 \sim 65\, km/Mpc.s$.
This result becomes quite interesting if we take into account some
recent considerations about a unification of cold dark matter and dark energy in
Chaplygin gas models \cite{bento}.
\par
The equation governing the evolution of our model is
\begin{equation}
\biggr(\frac{\dot a}{a}\biggl)^2 = \frac{8\pi G}{3}\biggr\{\frac{\rho_{m0}}{a^3} +
\sqrt{A + \frac{B}{a^6}}\biggl\} \quad \quad .
\end{equation}
It can be rewritten as
\begin{equation}
\biggr(\frac{\dot a}{a}\biggl)^2 = H_0\biggr\{\frac{\Omega_{m0}}{a^3} + \Omega_{c0}
\sqrt{\bar A + \frac{1 - \bar A}{a^6}}\biggl\}
\end{equation}
where $H_0$ is the Hubble parameter today and the scale factor was normalized to
unity today, $a_0 = 1$. We will consider only a flat spatial curvature section, which
seems to be favoured by observations \cite{charles}.
\par
The luminosity distance is obtained by a standard procedure \cite{coles}, using the equation
for the light trajectory in the above specified background,
and its definition,
\begin{equation}
D_L = \biggr(\frac{1}{4\pi}\frac{L}{l}\biggl)^{1/2}
\end{equation}
where $L$ is the absolute luminosity of the source, and $l$ is the luminosity
measured by
the observer.
This expression can be rewritten as
\begin{equation}
D_L = (1 + z)r \quad ,
\end{equation}
$r$ being the co-moving distance of the source. Taking into account the definitions
of absolute and apparent magnitudes in terms of the luminosity $L$ and $l$, $M$ and
$m$, respectively,
we finally obtain the relation
\begin{equation}
m - M = 5\log\biggr\{(1 + z)\int_0^z\frac{dz'}{[\Omega_{m0}(1 + z')^3 +
\Omega_{c0}\sqrt{\bar A + (1 - \bar A)(1 + z')^6}]|^{1/2}}\biggl\} \quad .
\end{equation}
\par
We proceed by fitting the supernova data using the model described above.
Essentially, we compute the quantity
\begin{equation}
\mu_0 = 5\log\biggr(\frac{D_L}{Mpc}\biggl) + 25 \quad,
\end{equation}
and compare the same quantity as obtained from observations.
The quality of the fitting is characterized by the parameter
\begin{equation}
\chi^2 = \sum_i\frac{[\mu_{0,i}^o - \mu_{0,i}^t]^2}{\sigma_{\mu_0,i}^2 + \sigma_{mz,i}^2}
\quad .
\end{equation}
In this expression, $\mu_{0,i}^o$ is the measured value, $\mu_{0,i}^t$ is the value
calculated through the model described above, $\sigma_{\mu_0,i}^2$ is the
measurement
error, $\sigma_{mz,i}^2$ is the dispersion in the distance modulus due to the
dispersion
in galaxy redshift caused by peculiar velocities. This quantity will be taken as
\begin{equation}
\sigma_{mz} = \frac{\partial \log D_L}{\partial z}\sigma_z \quad ,
\end{equation}
where, following \cite{riess,wang}, $\sigma_z = 200\,km/s$.
We evaluate, in fact, $\chi^2_\nu$, the estimated errors for degree of freedom.
\par
In order to compute $\chi^2_\nu$, we use data from \cite{riess,wang}.
The relevant data are listed in table $1$.
\begin{table}
\caption{The SN Ia data}
\vspace{0.3cm}
\begin{center}
\begin{tabular}{|c|c|c|||||c|c|c|}
\hline\hline
{\bf SN Ia}&{\bf z}&{\bf $\mu_{0}(\sigma_{\mu 0})$}&
{\bf SN Ia}&{\bf z}&{\bf $\mu_{0}(\sigma_{\mu 0})$} \\
&&&&&\\ \hline\hline
\bf {1992bo}&0.018&34.72(0.16)&\bf {1992br}&0.087&38.21(0.19) \\
&&&&&\\ \hline
\bf {1992bc}&0.020&34.87(0.11)&\bf {1992bs}&0.064&37.61(0.14) \\
&&&&&\\ \hline
\bf {1992aq}&0.111&38.41(0.15)&\bf {1993O}&0.052&37.03(0.12) \\
&&&&& \\ \hline
\bf {1992ae}&0.75&37.80(0.17)&\bf {1993ag}&0.050&36.80(0.17) \\
&&&&& \\ \hline
\bf {1992P}&0.026&35.76(0.13)&\bf {1996E}&0.43&42.03(0.22) \\
&&&&& \\ \hline
\bf {1990af}&0.050&36.53(0.15)&\bf {1996H}&0.62&43.01(0.15) \\
&&&&& \\ \hline
\bf {1992ag}&0.026&35.37(0.23)&\bf {1996I}&0.57&42.83(0.21) \\
&&&&&\\ \hline
\bf {1992al}&0.014&33.92(0.11)&\bf {1996J}&0.30&40.99(0.25) \\
&&&&&\\ \hline
\bf {1992bg}&0.035&36.26(0.21)&\bf {1996K}&0.38&42.21(0.18) \\
&&&&&\\ \hline
\bf {1992bh}&0.045&36.91(0.17)&\bf {1996U}&0.43&42.34(0.17) \\
&&&&&\\ \hline
\bf {1992bl}&0.043&36.26(0.15)&\bf {1997cl}&0.44&42.26(0.16) \\
&&&&&\\ \hline
\bf {1992bp}&0.080&37.75(0.13)&\bf {1997cj}&0.50&42.70(0.16) \\
&&&&&\\ \hline
\bf {1997ck}&0.97&44.30(0.19)&\bf {1995K}&0.48&42.49(0.17) \\
&&&&&\\ \hline
\end{tabular}
\end{center}
\end{table}
We compute $\chi^2_\nu$ varying $H_0$, $\Omega_{m0}$, $\Omega_{c0}$ and $\bar A$.
As an example, in tables $2$ and $3$ the values for $\chi^2_\nu$ are listed for the cases where
$H_0 = 65\,km/Mpc.s$ and $H_0 = 75\, km/Mpc.s$.
\begin{table}
\caption{Value of $\chi^2_\nu$ for the case where $H_0 = 65\,km/Mpc.s$}
\vspace{0.3cm}
\begin{tabular}{|l|c|c|c|c|c|c|c|c|c|c|}
\hline\hline
{~~~~~~~~~~~~~{$\bar A$}}&{\bf 0,1}&{\bf 0,2}&{\bf 0,3}&{\bf 0,4}
&{\bf 0,5}&{\bf 0,6}&{\bf 0,7}&{\bf 0,8}&{\bf 0,9}
&{\bf 1}\\ \bf $\Omega_{m0}$/$\Omega_{c0}$&&&&&&&&&&\\ \hline\hline
\bf {0.0/1.0}&4.689&4.226&3.761&3.297&2.838&2.391&1.974&1.619&1.412&1.749 \\
&&&&&&&&&&\\ \hline
\bf {0.1/0.9}&4.734&4.316&3.893&3.467&3.041&2.619&2.209&1.828&1.518&1.445 \\
&&&&&&&&&& \\ \hline
\bf {0.2/0.8}&4.780&4.406&4.026&3.642&3.253&2.861&2.470&2.087&1.729&1.458 \\
&&&&&&&&&& \\ \hline
\bf {0.3/0.7}&4.825&4.497&4.162&3.820&3.471&3.116&2.753&2.386&2.017&1.660 \\
&&&&&&&&&& \\ \hline
\bf {0.4/0.6}&4.871&4.588&4.298&4.001&3.696&3.381&3.056&2.718&2.364&1.986 \\
&&&&&&&&&& \\ \hline
\bf {0.5/0.5}&4.917&4.680&4.437&4.186&3.926&3.657&3.375&3.077&2.758&2.399 \\
&&&&&&&&&& \\ \hline
\bf {0.6/0.4}&4.963&4.773&4.576&4.373&4.162&3.941&3.708&3.459&3.188&2.874 \\
&&&&&&&&&&\\ \hline
\bf {0.7/0.3}&5.010&4.866&4.717&4.563&4.402&4.233&4.054&3.860&3.647&3.397 \\
&&&&&&&&&&\\ \hline
\bf {0.8/0.2}&5.055&4.959&4.860&4.756&4.647&4.532&4.410&4.277&4.129&3.955 \\
&&&&&&&&&&\\ \hline
\bf {0.9/0.1}&5.102&5.054&5.003&4.951&4.896&4.837&4.775&4.707&4.631&4.541 \\
&&&&&&&&&&\\ \hline
\bf {1.0/0.0}&5.148&5.148&5.148&5.148&5.148&5.148&5.148&5.148&5.148&5.148 \\
&&&&&&&&&&\\ \hline
\end{tabular}
\end{table}
\begin{table}
\caption{Value of $\chi^2_\nu$ for the case where $H_0 = 75\,km/Mpc.s$}
\vspace{0.3cm}
\begin{tabular}{|l|c|c|c|c|c|c|c|c|c|c|}
\hline\hline
{~~~~~~~~~~~~~{$\bar A$}}&{\bf 0,1}&{\bf 0,2}&{\bf 0,3}&{\bf 0,4}
&{\bf 0,5}&{\bf 0,6}&{\bf 0,7}&{\bf 0,8}&{\bf 0,9}
&{\bf 1}\\ \bf $\Omega_{m0}$/$\Omega_{c0}$&&&&&&&&&&\\ \hline\hline
\bf {0.0/1.0}&12.525&11.674&10.792&9.876&8.922&7.926&6.883&5.792&4.664&3.635 \\
&&&&&&&&&&\\ \hline
\bf {0.1/0.9}&12.608&11.841&11.046&10.220&9.358&8.454&7.502&6.494&5.418&4.279 \\
&&&&&&&&&& \\ \hline
\bf {0.2/0.8}&12.690&12.008&11.301&10.565&9.796&8.988&8.132&7.218&6.225&5.110 \\
&&&&&&&&&& \\ \hline
\bf {0.3/0.7}&12.772&12.176&11.556&10.911&10.236&9.526&8.771&7.960&7.069&6.042 \\
&&&&&&&&&& \\ \hline
\bf {0.4/0.6}&12.855&12.343&11.812&11.258&10.678&10.067&9.416&8.714&7.938&7.032 \\
&&&&&&&&&& \\ \hline
\bf {0.5/0.5}&12.937&12.511&12.068&11.606&11.122&10.610&10.066&9.477&8.824&8.056 \\
&&&&&&&&&& \\ \hline
\bf {0.6/0.4}&13.020&12.678&12.324&11.954&11.566&11.156&10.719&10.246&9.720&9.102 \\
&&&&&&&&&&\\ \hline
\bf {0.7/0.3}&13.102&12.846&12.580&12.302&12.011&11.703&11.375&11.019&10.624&10.159 \\
&&&&&&&&&&\\ \hline
\bf {0.8/0.2}&13.184&13.014&12.836&12.651&12.457&12.251&12.032&11.795&11.531&11.222 \\
&&&&&&&&&&\\ \hline
\bf {0.9/0.1}&13.267&13.181&13.093&13.000&12.903&12.800&12.691&12.572&12.440&12.286 \\
&&&&&&&&&&\\ \hline
\bf {1.0/0.0}&13.349&13.349&13.349&13.349&13.349&13.349&13.349&13.349&13.349&13.349 \\
&&&&&&&&&&\\ \hline
\end{tabular}
\end{table}
In general, the best results are obtained, in each case when $\Omega_{m0} = 0$,
$\Omega_{c0} = 1$ and $\bar A = 1$. These cases represent a pure cosmological contant
model. However, the best fitting is in fact obtained when $\Omega_{m0} = 0$,
$\Omega_{c0} = 1$, $H_0 = 65\,km/Mpc.s$ and $\bar A \sim 0.92$. This case represents
a universe containing just the Chaplygin gas, which exhibits a behaviour close
to a cosmological constant. In this case, the universe begins to accelerate
at $z \sim 0.7$. In figure $1$ the fitting for this case is exhibited.
\begin{figure}
\centering
\psfig{file=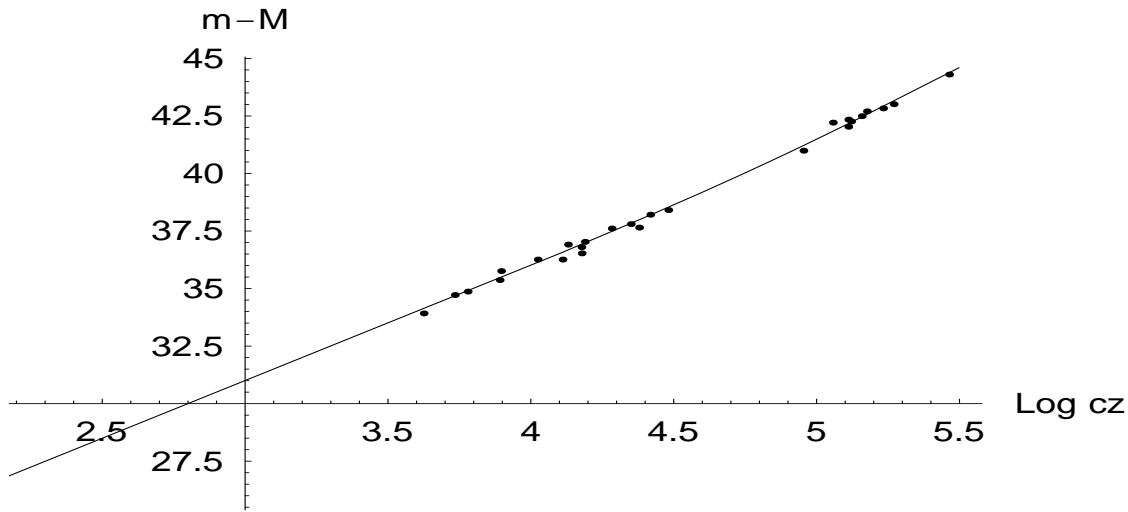,width=15cm,height=7cm}
\caption{$\Omega_{m0} = 0$,
$\Omega_{c0} = 1$, $H_0 = 65\,km/Mpc.s$ and $\bar A \sim 0.92$}
\label{figure 1}
\end{figure}
\par
The fact that the best fitting is achieved by a model with the Chaplygin gas as
the only component of the matter content of the universe may be seen as a negative
feature of the results discussed above. However, some comments must be made
about this point. First, we have neglected the baryon content, which must
contribute with a factor $\Omega \sim 0.2\,h^{-1}$, where $h = H_0/(100\,km/Mpc.s)$,
as deduced from the primordial nucleosynthesis and from the spectrum of the anisotropy of
the cosmic microwave background radiation. But, the introducing of the baryon component
does not change substantially the results above. Second, there is evidence that
the Chaplygin gas may unify the cold dark matter and dark energy scenarios
\cite{bento}, in the sense that it can behave as cold dark matter
at small scales and
as dark energy at large scales. Hence, our results may be an indication that
such a
unification of dark matter and dark energy through the Chaplygin gas must be taken
more seriously. In order to confirm this, the analysis of the anisotropy of
cosmic microwave background radiation in this scenario may be performed. We hope
to present this result in the future.
\newline
\vspace{0.5cm}
\newline
\noindent
{\bf Acknowledgements:}  We thank Kirill Bronniko and Fl\'avio Gimenes Alvarenga
for careful reading of the manuscript. This work has receveid partial
financial supporting from CNPq (Brazil).

\end{document}